\documentclass[preprint,aps,12pt,showpacs,nofootinbib,tightenlines]{revtex4}
\usepackage{amsmath}
\usepackage{amssymb}
\usepackage{epsfig}
\usepackage{graphicx}
\begin{document}
%%%%%%%%%%%%%%%%%%%%%%%%%%%%%%%%%%%%%%%%%%%%%
%%%%%%%%%%%%%%%%%%%%%%%%%%%%%%%%%%%%%%%%%%%%%

\newcommand{\beq}{\begin{eqnarray}}
\newcommand{\eeq}{\end{eqnarray}}
\newcommand{\non}{\nonumber\\ }

\newcommand{\mb}{m_B }
\newcommand{\mtp}{m_{t'}}
\newcommand{\mw}{m_W }
\newcommand{\im}{{\rm Im} }

%%---------------------------------------------------------

\def \cpc{ {\bf Chin. Phys. C} }
\def \ctp{ {\bf Commun.Theor.Phys. } }
\def \epjc{{\bf Eur.Phys.J. C} }
\def \jpg{ {\bf J.Phys. G} }
\def \npb{ {\bf Nucl.Phys. B} }
\def \plb{ {\bf Phys.Lett. B} }
\def \pr{  {\bf Phys. Rep.} }
\def \prd{ {\bf Phys.Rev. D} }
\def \prl{ {\bf Phys.Rev.Lett.}  }
\def \ptp{ {\bf Prog. Theor. Phys. }  }
\def \rmp{ {\bf Rev.Mod.Phys. }  }
\def \zpc{ {\bf Z.Phys.C}  }
\def \jhep{ {\bf J. High Energy Phys.}  }
\def \epl{ {\bf Europhys. Lett.} }
%%%%%%%%%%%%%%%%%%%%%%%%%%%%%%%%%%%%%%%%%%%%%%%%%%%%
%%
\title{Production and decays of a light $\phi^0$ in the LRTH model
under the LHC Higgs data}
\author{Yao-Bei Liu$^{1,2}$, Zhen-Jun Xiao$^{1}$\footnote{Electronic address: xiaozhenjun@njnu.edu.cn}}
\affiliation{1. Department of Physics and Institute of Theoretical
Physics, Nanjing Normal University, Nanjing 210023,
P.R.China \\
2. Henan Institute of Science and Technology, Xinxiang
453003, P.R.China} %%

%%\date{\today}
\begin{abstract}
In this paper we study the production and decays of a light
pseudoscalar boson $\phi^0$ with $m_{\phi^0} \leq m_h/2$ appeared in the
left-right twin Higgs (LRTH) model, and explore its phenomenological
consequences when the latest LHC Higgs data are taken into account.
We found that
(a) the decay rate $Br(h\to \phi^0\phi^0)$ can be as large as $80\%$ and can suppress significantly the
visible $\gamma\gamma$ signal rate, but the latest LHC Higgs data put a strong constraint on it:
$Br(h \to \phi^0\phi^0) \leq 30\%$ at $3\sigma$ level;
(b) the $p$-value of the LRTH model is around $0.6$, smaller than that of the SM in most of the parameter
space and approaches the SM value $0.8$ for a sufficiently large $f$ parameter;
(c) the neutral pseudoscalar $\phi^0$ dominantly decay into $b\bar{b}$ and the decay rate $Br(\phi^0\to b\bar{b})$ can be
larger than $80\%$ for $m_{\phi^0}\leq 60$ GeV, and the second main decay mode is $\phi^0\to \tau^{+}\tau^{-}$ with
a branching ratio about $14\%$;
and (d) at the future $e^-e^+$ collider with $\sqrt{s}=250$ GeV, the processes $e^{+}e^{-}\to Zh\to
Z(\phi^0\phi^0)\to Z(4b,2b2\tau)$ are promising for discovering such a light pseudoscalar $\phi^0$.
\end{abstract}

\maketitle

%%====================================================================
\newpage
\section{Introduction}

The discovery of a neutral Higgs boson with a mass around 125 GeV at
CERN's Large Hadron Collider (LHC) has been confirmed by
the ATLAS and CMS collaborations \cite{1,2,1b,2b,4,5}, which heralds the
beginning of a new era of Higgs physics. So far the observed signal
strengths, albeit with large experimental uncertainties, consistent
with the Standard Model (SM) predictions \cite{4,5}.
However, the SM suffers from the so-called gauge hierarchy problem and
cannot provide a dark matter candidate.

During past three decades, many new physics (NP) models beyond the SM have been  constructed by extending the
Higgs sector in the SM, such as the Supersymmetric (SUSY) models \cite{susy},
large extra-dimensions \cite{led}, two-Higgs doublet models (2HDM) \cite{2hd},
and little Higgs models \cite{little,little1,little2} etc.
Very recently, the twin Higgs models have been proposed  \cite{ly-1,ly-11,ly-12,ly-13}  as
a solution to the little hierarchy problem.
Here we focus on the left-right twin Higgs (LRTH) model which is implemented with the
discrete symmetry being identified with left-right symmetry \cite{ly-2,ly-3}.
In the LRTH model, several physical Higgs bosons are still left after the spontaneous symmetry breaking.
Another additional discrete symmetry is introduced under an odd $SU(2)_{L}$ doublet $\hat{h}$
while the other fields are even.
The lightest particle in the neutral components $\hat{h}^{0}_{2}$ is stable and can be a candidate
for weakly interacting massive particle (WIMP) dark matter.
Besides $\hat{h}^{0}_{2}$, the LRTH model predicts the SM-like Higgs boson $h$ and other three
scalars: $\phi^{0}$ and $\phi^{\pm}$.

 The neutral $\phi^{0}$ is a pseudoscalar and thus there are
no $\phi^{0}W^{+}W^{-}$ and $\phi^{0}ZZ$ couplings at the tree level.

, which makes the $\phi^{0}$
rather special.
The particle spectrum and collider signatures of the LRTH model have been
widely studied, for example, in Refs.~ \cite{Hock,dong,sf,wl,wl1,prd81,prd77,prd79,prd79cha}.

In a recent paper \cite{liu-1311}, we studied the properties of the SM-like Higgs boson $h$,
calculated the new physics contributions to the decays $h\to (\gamma\gamma, Z\gamma,
\tau\tau, WW^*,ZZ^*,\tau\tau)$
in the LRTH model, performed a globe fit to the current LHC Higgs data, and found that
all the signal rates are suppressed when NP contributions are taken into account, while
the LRTH prediction for $R_{\gamma\gamma}$ agrees well with the CMS measurement
$R_{\gamma\gamma}=0.77\pm 0.27$ at $1\sigma$ level.
In this paper, we will study the production and decays of the light
pseudo-scalar $\phi^0$ and to
draw the possible constraints from currently available LHC Higgs data.
If this neutral $\phi^0$ were lighter than half of the SM-like Higgs
boson $h$, i.e. $m_{\phi^0}< m_{h}/2$,
the new decay channel $h \to \phi^0 \phi^0$ will be opened with a sizable branching ratio.
Because the 125 GeV SM Higgs decay width is small ( the measured value is about 4 MeV),
such an exotic decay mode can suppress greatly the visible signals of $h$ and would
have important phenomenological consequences \cite{ex1,ex2,ex3,ex4}.
We know that the current bound on the branching ratios to exotic states is
still weak: a branching
fraction as large as $\sim 60\%$ is allowed at the $2 \sigma$ C.L. \cite{4,5}.
If SM couplings are assumed, the universal Higgs fits constrain the invisible
branching fraction to be less than
$25\%$ at $95\%$ C.L. \cite{cons1,cons11,cons12}, which still leaves appreciable scope for
such an exotic decay mode. Thus, we will investigate the constrains of the
latest LHC Higgs data on the properties of such a light
pseudoscalar $\phi^0$ in the LRTH model.
We will also study the possibility of detecting such a light boson $\phi^0$
at high energy colliders.

This paper is organized as follows. In the next section, we briefly review the LRTH model
and study the possible
decay modes for a light pseudoscalar boson $\phi^0$. In Sec. III, we investigate the decay
branching ratios of $h\to \phi^0\phi^0$ and perform a fit using the latest LHC
Higgs data. We study the possibility of detecting such a light pseudoscalar
at the LHC experiments in section IV. Finally, we present our conclusion in Sec.V.

\section{The left-right twin Higgs model}\label{sec:intro}

\subsection{ Outline of the LRTH model}

This model is based on the global $U(4)\times U(4)$ symmetry with a locally gauged
subgroup $SU(2)_{L}\times SU(2)_{R}\times
U(1)_{B-L}$ \cite{ly-1,ly-11,ly-12,ly-13,ly-2}.
The twin symmetry is identified as the left-right
symmetry which interchanges L and R, implying that the gauge
couplings of $SU(2)_{L}$ and $SU(2)_{R}$ are identical $(g_{2L}=g_{2R}=g_{2})$.
Two Higgs fields, $H$ and $\hat{H}$, are introduced
and each transforms as $(4,1)$ and $(1,4)$, respectively.
They are written as \beq H=\left( \begin{array}{c} H_{L}\\ H_{R} \\
\end{array}  \right)\,, \qquad
\hat{H}=\left( \begin{array}{c} \hat{H}_{L}\\ \hat{H}_{R} \\
\end{array}  \right)\,,
\eeq where $H_{L,R}$ and $\hat{H}_{L,R}$ are two component objects
which are charged under the $SU(2)_{L}\times SU(2)_{R}\times
U(1)_{B-L}$ as \beq H_{L}~and~ \hat{H}_{L}: (2, 1, 1),~~~~~~~~H_{R}~
and~ \hat{H}_{R}: (1, 2, 1). \eeq The global $U(4)_{1}(U(4)_{2})$
symmetry is spontaneously broken down to its subgroup
$U(3)_{1}(U(3)_{2})$ with non-zero vacuum expectation values(VEV):
\beq
<H>=(0,0,0,f)^{\rm T}, \quad  <\hat{H}>=(0,0,0,\hat{f})^{\rm T}.
\eeq
Each spontaneously symmetry breaking yields seven Nambu-Goldstone bosons,
which can be parameterized as
\beq
H=fe^{\frac{i\pi}{f}}\left( \begin{array}{c} 0\\ 0\\ 0\\ 1\\
\end{array}
\right),~~~~~~~~~~~~{\rm with\ }
 \pi=\left( \begin{array}{cccc} -N/2&0&   0&    h_{1}\\
0&  -N/2&   0&   h_{2}\\ 0&  0 &  -N/2 &  C\\
h_{1}^{*}&   h_{2}^{*}&   C^{*}&   3N/2\\
\end{array} \right),
\eeq
where $\pi$ are the corresponding Goldstone fields. N
is a neutral real pseudoscalar, $C$ and $C^{*}$ are a pair of
charged complex scalar fields. $(h_{1},h_{2})$ is the SM $SU(2)_{L}$
Higgs doublet. Accordingly, $\hat{H}$ is parametrized in the same
way by its own Goldstone boson matrix $\hat{\pi}$, which contains
$\hat{N}$, $\hat{C}$ and
$\hat{h}=(\hat{h}_{1}^{+},\hat{h}_{2}^{0})$.

The original gauge symmetry $SU(2)_{L}\times SU(2)_{R}\times U(1)_{B-L}$ is
broken down to the SM $U(1)_{Y}$,
six out of the 14 Goldstone bosons are respectively eaten by the SM gauge bosons
$W^{\pm}$ and $Z$, and additional gauge bosons $W^{\pm}_{H}$, and $Z_{H}$
with masses of TeV order.
Then we are left with the SM-like physical Higgs boson $h$,  one neutral
pseudoscalar $\phi^{0}$,
a pair of charged scalar $\phi^{\pm}$, and an odd $SU(2)_{L}$ doublet
$\hat{h}=(\hat{h}_{1}^{+},\hat{h}_{2}^{0})$
which only couples to the gauge boson sector.
The lightest particle in $\hat{h}$ is stable and thus can be a candidate for WIMP dark matter,
which have been studied for example in Refs. \cite{sf,wl}.

The covariant kinetic terms of Higgs fields can be written as
\cite{Hock}
\begin{eqnarray}
    {\cal L}_H &=& (D_{\mu}H)^{\dagger}D^{\mu}H +
(D_{\mu}\hat{H})^{\dagger}D^{\mu}\hat{H},
\end{eqnarray}
where the covariant derivative is $ D^{\mu} = \partial^{\mu} - i g_2 W_2^{\mu}
-ig_1n_{B-L}W_{B-L}^{\mu}$, and
\begin{equation}
 W_2 =\frac{1}{2}\left(%
\begin{array}{cccc}
  W^0_L & \sqrt{2}W^+_L & 0 & 0 \\
  \sqrt{2}W^-_L & -W^0_L & 0 & 0 \\
   0 & 0 & W^0_R & \sqrt{2}W^+_R \\
   0 & 0 & \sqrt{2}W^-_R & -W^0_R \\
\end{array}%
\right),\ \ \
    W_{B-L} =\frac{W_1}{2}
\left(%
\begin{array}{cccc}
  1 & 0& 0 & 0 \\
 0 & 1 & 0 & 0 \\
   0 & 0 & 1& 0 \\
   0 & 0 & 0 & 1 \\
\end{array}%
\right),
\end{equation}
where $g_1$ and $g_2$ are the gauge couplings for ${\rm U}(1)_{B-L}$
and ${\rm SU}(2)_{L,R}$, $n_{B-L}=1$ is the charge of the field under ${\rm U}(1)_{B-L}$.

In the LRTH model, a pair of vector-like quarks $(U_L,U_R)$ are introduced to cancel the
one-loop quadratic divergence of Higgs mass induced by the top
quark. The relevant Lagrangian can be written as \cite{Hock}
\begin{equation}
{\cal L}_{t}=y_L\bar{Q}_{L3}\tau_2 H_L^*U_R
  +y_R\bar{Q}_{R3}\tau_2H_R^*U_L - M\bar{U}_LU_R + h.c.
\label{yukawatop}
\end{equation}
where $Q_{L3} = -i(u_{L3},d_{L3})^{\rm T}$ and $Q_{R3} = (u_{R3},d_{R3})^{\rm T}$.

 The details of the LRTH model as well as the
particle spectrum, Feynman rules, and some phenomenology analysis
have been given for example in Ref.~\cite{Hock}. Here we will focus on the properties of the
light pseudoscalar $\phi^0$.

\subsection{The mass and decay of the light scalar $\phi^0$}

 In the LRTH model, the soft left-right symmetry breaking
terms, so called $\mu-$term, can generate mass for the light $\phi^0$:
\beq
V_{\mu}=-\mu_{r}^{2}(H_{R}^{\dagger}\hat{H}_{R}+h.c.)+\hat{\mu}^{2}H_{L}^{\dagger}\hat{H}_{L}.
\eeq
The mass of $\phi^0$ and new scalar self-interactions are given by \cite{Hock}
\begin{eqnarray}
m^{2}_{\phi^0}&=&\frac{\mu_{r}^{2}f\hat{f}}{\hat{f}^{2}+f^{2}\cos^{2}x}\non
&& \cdot \left \{\frac{\hat{f}^{2}\left[ \cos
x+\frac{\sin x}{x}(3+x^{2}) \right]}{f^{2}\left (\cos x+\frac{\sin x}{x} \right )^{2}}+2\cos
x+\frac{f^{2}\cos^{2}x(1+\cos x)}{2\hat{f}^{2}} \right\},\label{eq:mphi2}\\
h\phi^0\phi^0 &:& \frac{v m_h^2}{54f^2}\cdot
\left [ 11 + 15\left ( 1-\frac{2m^2_{\phi^0}}{m_h^2} \right)\right ],
 \end{eqnarray}
where $x=v/(\sqrt{2}f)$ and $v=246$ GeV is the electroweak scale. Once $f$ is fixed, the scale $\hat{f}$ can be
determined from the electroweak symmetry breaking condition.
In general $\hat{f}$ is larger $f$ about 5 times or more \cite{Hock,dong} and here we set $\hat{f}=5f$ as a rough estimate.

%%%%%%%%%%%%%%%%%%%%%%%%%%%%%%%%%%%%%%%%%%%%%%%%%%%%
\begin{figure}[tbh]
\begin{center}
\vspace{1cm} \centerline{\epsfxsize=8cm\epsffile{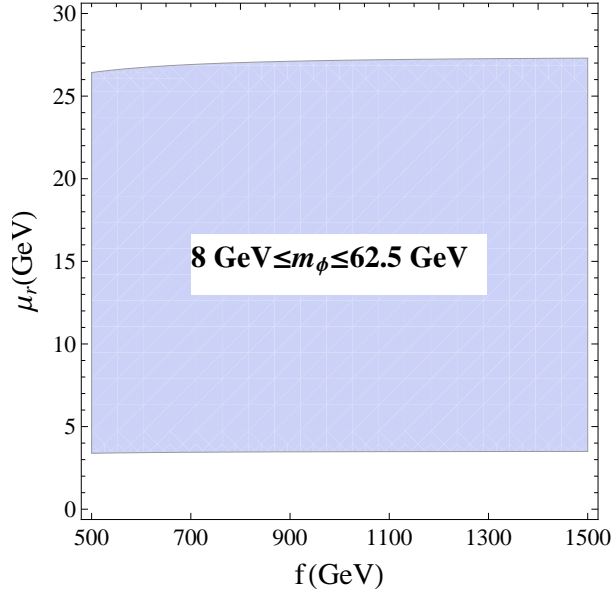}}
\caption{The upper constraint $m_{\phi^0}\leq 62.5$ GeV, according to the theoretical
relation as given
in Eq.~(\ref{eq:mphi2}).}
\label{fig:fig1}
\end{center}
\end{figure}

From the expression of $m_{\phi^0}^2$ in Eq.~(\ref{eq:mphi2}), one can see that the value of
$m_{\phi^0}^2$ depend on two parameters $\mu_r$ and $f$.
The value of $\mu_{r}$ cannot be too large, since the fine-tuning of the SM-like
Higgs boson mass $m_h$ will become severe for larger $\mu_r$ \cite{Hock}.
Assuming $4\leq \mu_r \leq27$ GeV and $500 \leq f\leq 1500$ GeV, one finds
the upper constraint on $m_{\phi^0}$:
$m_{\phi^0} \leq 62.5$ GeV. In other words, the new decay channel $h\to
\phi^0\phi^0$ can be opened in the
parameter space of the LRTH model considered here.
The lower limit of $m_{\phi^0}$, say $m_{\phi^0} > 7$ GeV, comes from the
non-observation of the decay
$\Upsilon\to \gamma + X_{0}$ \cite{cons2,cons21}.
As illustrated in Fig.~\ref{fig:fig1}, the upper limit $ m_{\phi^{0}} \leq
62.5$ GeV are guaranteed
when the values of $(f,\mu_r)$ are in the light-dark region in the $f-\mu_r$ plane.
The rare decays of $Z\to f\bar{f}\phi^{0}$ and $Z\to \phi^{0}\gamma$ have
been studied in Ref.~\cite{wanglei}.

%%----------------------------------  decays of $\phi^0$

In the LRTH model, the decays $\phi^0\to gg, \gamma\gamma$ are mediated
by the one loop Feynman diagrams
involving the top quark and the new heavy quark $T$. The leading order decay
widths can be written as \cite{zz}
\beq
 \Gamma(\phi^0\to gg)&=&\frac{\sqrt{2}G_{F}\alpha_{s}^{2}m_{\phi^0}^{3}}{32\pi^{3}}
 \left |-\frac{1}{2}F_{1/2}(\tau_{t})y_{t}-
 \frac{1}{2}F_{1/2}(\tau_{T})y_{T} \right |^{2},\\
 \Gamma(\phi^0\to  \gamma\gamma)&=&\frac{\sqrt{2}G_{F}\alpha_{e}^{2}m_{\phi^0}^{3}}{256\pi^{3}}
\left  |\frac{4}{3}F_{1/2}(\tau_{t})y_{t}+\frac{4}{3}F_{1/2}(\tau_{T})y_{T}\right |^{2},
\eeq
where $F_{1/2}=-2\tau[1+(1-\tau)f(\tau)]$  with $f(\tau)=[\sin^{-1}(1/\sqrt{\tau})]^{2}$
and $\tau_{t}=4m_t^2/m_{\phi^0}^2$ , $\tau_{T}=4m_T^2/m_{\phi^0}^2$.
The explicit expressions of the relevant couplings  $y_{t}$ and $y_{T}$ are of the form
\beq
y_{t}= S_{L}S_{R},\quad y_{T}&=& \frac{m_{t}}{m_{T}}C_{L}C_{R},
\eeq
where the mixing angles $S_{L,R}$ and $C_{L,R}$ are
\beq
S_{L}&=& \frac{1}{\sqrt{2}}\sqrt{1-(y^{2}f^{2}\cos2x+M^{2})/N_{t}},
\quad C_{L}=\sqrt{1-S_L^2}, \label{eq:sl1}\\
S_{R}&=& \frac{1}{\sqrt{2}}\sqrt{1-(y^{2}f^{2}\cos2x-M^{2})/N_{t}},
\quad C_{R}=\sqrt{1-S_{R}^2},\label{eq:sr1}
\eeq
with
\beq
N_{t}=\sqrt{(M^{2}+y^{2}f^{2})^{2}-y^{4}f^{4}\sin^{2}2x},\label{eq:nt1}
\eeq
where $x=v/(\sqrt{2}f)$. The mass of the top quark and new heavy $T$-quark are
therefore can also be written as \cite{Hock}
\beq
m_{t}^{2}= \frac{1}{2}(M^{2}+y^{2}f^{2}-N_{t}),\quad
m_{T}^{2}= \frac{1}{2}(M^{2}+y^{2}f^{2}+N_{t}).
\label{eq:mt2}
\eeq
The parameter $y$ in Eqs.~(\ref{eq:sl1}-\ref{eq:mt2}) denotes the top quark
Yukawa coupling, and can be
determined by fitting the measured value of $m_t$ according to Eq.~(\ref{eq:mt2})
for given values of the new physics parameters $f$ and $M$.

For $\phi^{0}\to f_i\bar{f}_i$ decays with  $f_i$ the leptons and/or light quarks,
the decay width can be written as:
\beq
\Gamma(\phi^{0}\to f_i\bar{f}_i)=\frac{N_{C}G_{F}v^{2}m^{2}_{i}
m_{\phi^{0}}}{8\sqrt{2}\pi f^{2}}(1-x_i)^{3/2},
\eeq
where $x_i=4m_{i}^{2}/m_{\phi^{0}}^{2}$, $N_{c}=3(1)$ for $f_{i}$ being a quark (lepton).
It is easy to see that the decays of $\phi^{0}$ to those light final state fermions,
such as $f_i=(e,\mu, u,d,s)$,  are strongly suppressed  due to the severe
helicity  suppression ( $\propto m_i^2$), and therefore can be neglected safely.

%%-------------------------------------------------------

%%%%%%%%%%%%%%%%%%%%%%%%%%%%%%%%%%%%%%%%%%%%%%%%%%
\begin{figure}[thb]
\begin{center}
\vspace{-0.5cm} \centerline{\epsfxsize=8cm\epsffile{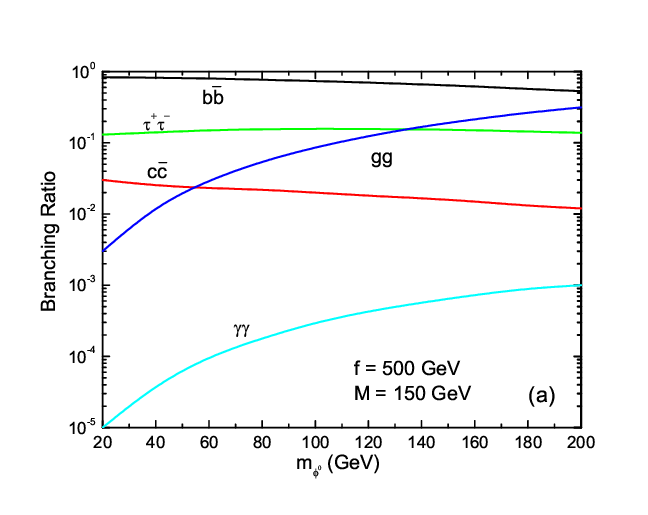}
\hspace{-0.5cm}\epsfxsize=8cm\epsffile{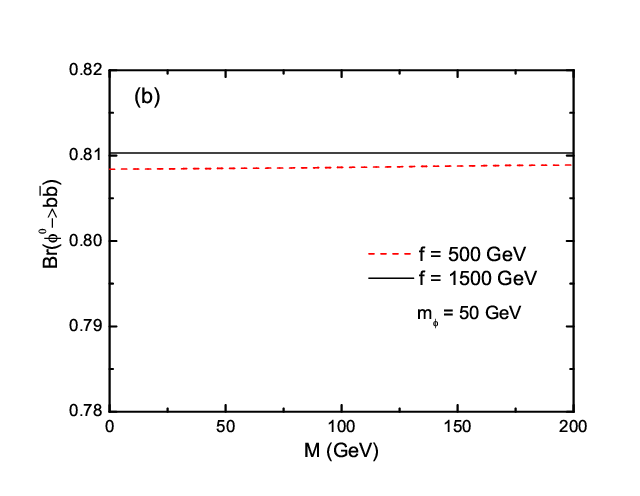}}
\caption{(a) the branching ratios of the considered $\phi^0$ decays as a
function of $m_{\phi^0}$ for given values of $f=500$ GeV and $M=150$ GeV.
(b) the branching ratio of the dominant $\phi^0\to b\bar{b}$ decay versus
$M$ for fixed $m_{\phi^{0}}=50$ GeV and
$f=500, 1500$ GeV.} \label{fig:fig2}
\end{center}
\end{figure}
%%%%%%%%%%%%%%%%%%%%%%%%%%%%%%%%%%%%%%%%%%%%%%%%%%

In the LRTH model, consequently, the five major decay modes of $\phi^0$ are
$\phi^0\to b\bar{b}, c\bar{c}$, $\tau^{+}\tau^{-}$, $gg$ and $\gamma\gamma$.
The $m_{\phi^0}$-dependence of the branching ratios, assuming $f=500$ GeV and
$M=150$ GeV, are illustrated in Fig.2a. The Fig.~2b shows the $M$-dependence of the
branching ratio of the dominant $\phi^0\to b\bar{b}$ decay for fixed $f=500$ and $1500$ GeV.
For $\phi^0\to f\bar{f}$ decays, furthermore, their decay rates in the
LRTH model are strongly suppressed by a factor of $v^{2}/(2f^{2}) \leq 0.12$
when compared with those of the SM Higgs boson decays $H\to f\bar{f}$.
From Fig.~2 one can see that:
\begin{enumerate}
\item
The dominant decay mode of the light pseudoscalar $\phi^0$ is $\phi^0\to b\bar{b}$.
In the considered region of $500 GeV \leq f\leq 1500 GeV$,
the value of the branching ratio $Br(\phi^0\to b\bar{b})$
is about $81\%$ for $m_{\phi^{0}}=50$ GeV, and
has a rather weak dependence on the variations of the parameters $f$ and $M$.

\item
The partial width into $c\bar{c}$ is smaller than that into $\tau^{+}\tau^{-}$, this is
because we use
the running mass of the quarks evaluated at the scale $m_{\phi^{0}}$ to calculate the Yukawa
coupling. In the allowed parameter spaces, $Br(\phi^0\to \tau^{+}\tau^{-})\simeq 14\% $.

\item
$Br(\phi^0\to gg$) becomes large along with the increase of $m_{\phi^0}$,
which can reach $30\%$ for $m_{\phi^0}=200$ GeV. This is due to the enhancement
from the contribution of heavy $T$-quark, which is non-decoupled in the triangle loops.

\item
The values of $Br(\phi^0\to \gamma\gamma$) is very small:
at the level of $10^{-5}$ to $10^{-3}$ in most of the parameter
space. This is due to the absence of the coupling between
$\phi^0$ and the charged gauge bosons.

\end{enumerate}

\section{Effects of a light $\phi^0$ and the LHC Higgs data}

In our calculations, we take the SM-like Higgs mass as $m_{h}=125.5$ GeV.
The SM input parameters relevant in our study are taken from \cite{data}.
The free parameters in the LRTH model relevant for this work
are $f$, $M$ and $m_{\phi^0}$.
Following Ref.~\cite{Hock}, we here also assume that the values of the free
parameters $f$ and $M$ are in the ranges of
\beq
500 \leq f \leq 1500 {\rm GeV}, \quad 0 \leq M \leq 150 {\rm GeV}.
\eeq
while $8\leq m_{\phi^0} \leq 62.5$ GeV according to the analysis in previous section.

\subsection{The $h\to \phi^0 \phi^0$ decay}

For $m_{h}\geq 2m_{\phi^0}$,  the new decay channel $h\to \phi^0\phi^0$ will  open and the partial decay
width is given by
\beq
\Gamma(h\to
\phi^0\phi^0)=\frac{g^{2}_{h\phi^0\phi^0}}{8\pi
m_{h}}\sqrt{1-\frac{4m_{\phi}^{2}}{m_{h}^{2}}},
\eeq
where $g_{h\phi^0\phi^0}$ is the coupling of $h\phi^0\phi^0$ vertex.
The open of this new decay mode, consequently, can suppress greatly the visible
signals of the boson $h$ at the LHC.
Thus, the major decay modes of the SM-like Higgs  boson $h$ in the LRTH model become now:
\beq
h\to \phi^0\phi^0, {\rm \ \ and \ \ } h\to f\bar{f} (f=b,c,\tau), VV^{*}(V=W, Z),
gg, \gamma\gamma,
\eeq
where $W^{*}/Z^{*}$ denoting the off-shell charged or neutral electroweak
gauge bosons. The branching ratio of $h\to \phi^0\phi^0$ can be written as
\beq
Br(h\to \phi^0\phi^0)=\frac{\Gamma(h\to
\phi^0\phi^0)}{\Gamma_{\rm LRTH}(h)+\Gamma(h\to
\phi^0\phi^0)},
\eeq
where $\Gamma_{\rm LRTH}(h)$ denotes the total decay width of SM-like Higgs boson $h$
for $m_{\phi^0} > m_{h}/2$ in the LRTH model, which has been studied in
Refs.\cite{liu-plb,liu-1311}.

%%%%%%%%%%%%%%%%%%%%%%%%%%%%%%%%%%%%%%%%%%%%%%%%%%
\begin{figure}[thb]
\begin{center}
\vspace{-0.5cm}
\centerline{\epsfxsize=8cm\epsffile{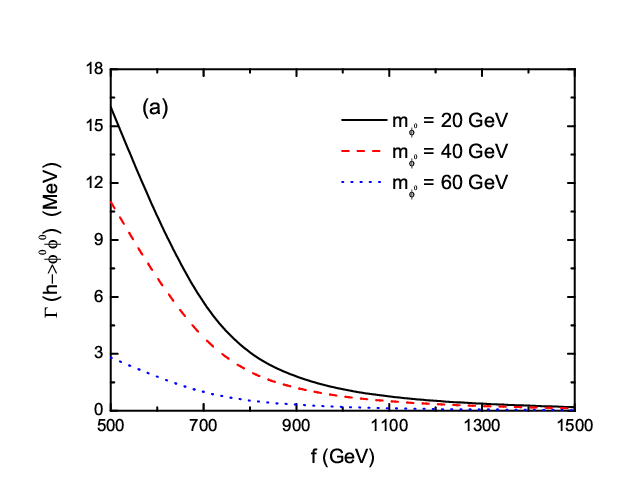}
\hspace{-0.5cm}\epsfxsize=8cm\epsffile{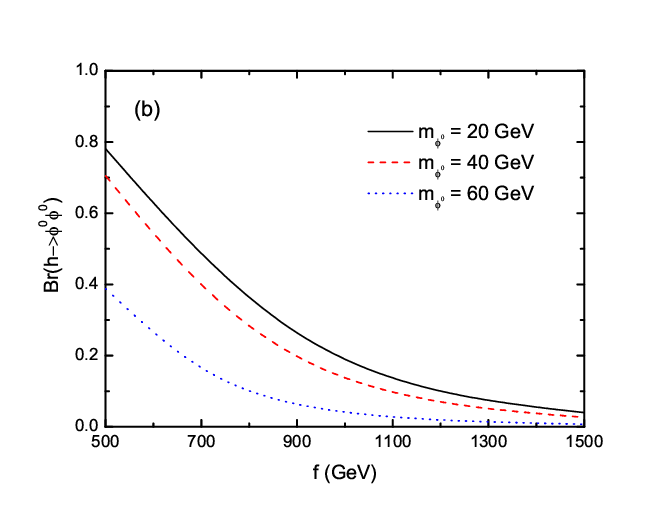}}
\caption{ The $f$-dependence of $\Gamma(h\to \phi^0\phi^0)$ (left)
and $Br(h\to \phi^0\phi^0)$ (right) for $M=150$ GeV and
three typical values of $m_{\phi^0}=20,40$ and $60$ GeV. }
\label{fig:fig3}
\end{center}
\end{figure}

In Fig.~3 we show the $f$-dependence of the decay width  $\Gamma(h\to \phi^0\phi^0)$
and the branching ratio  $Br(h\to \phi^0\phi^0)$  for $ M=150$ GeV and
three typical values of $m_{\phi^0}$:  $m_{\phi^0}= 40\pm 20$ GeV.
One can see that both the decay width and decay rates for $h\to \phi^0\phi^0$ decay
becomes smaller rapidly along with the increase of the parameter $f$.
This is because the couplings of $h\phi^0\phi^0$ is proportional to the suppression
factor of $(v/f)^2$. For $m_{\phi}=40$ GeV,  we find $2\% \leq Br(h\to \phi^0\phi^0)
\leq 70\%$ for $500 \leq f\leq 1500$ GeV.
For the case of $f=500$ GeV, the decay width $\Gamma(h\to \phi^0\phi^0)$
can be as large as 16 MeV and thus can suppress greatly the branching ratios
for other decay modes of the SM-like Higgs boson $h$: such
as the phenomenologically very interesting $h\to \gamma\gamma$ decay.

\subsection{$h\to \gamma\gamma$ decay in the LRTH model }

For the SM Higgs diphoton decay, the measured signal strength
as reported by ATLAS \cite{4} and CMS collaboration \cite{5}
are rather different,
\beq
R_{\gamma\gamma}=\frac{\sigma(H \to \gamma\gamma)}{\sigma^{\rm SM}(H\to \gamma \gamma)}
= \left \{ \begin{array}{ll} 1.55 ^{+0.33}_{-0.28},& {\rm ATLAS; }\\
0.77\pm 0.27,& {\rm CMS.}\\ \end{array} \right.
\eeq
but these results are still consistent with the SM expectation within 2$\sigma$ level
due to rather large errors.
If the excess (deficit) seen by ATLAS (CMS) were eventually confirmed by the
near future LHC measurements, the extra NP contributions would be help to
understand such excess or deficit \cite{np1,np2,np3,np4,np5}.

At the LHC, the Higgs single production is dominated by the gluon-gluon fusion (ggF) process.
The hadronic production cross section $\sigma(gg\to h)$ has a strong
correlation with the decay width $\Gamma(h\to gg)$. Other
main production processes of the Higgs boson include vector-boson
fusion (VBF), associated production with a $W/Z$ boson (VH)
and associated production with a $t\bar{t}$ pair (ttH).
For $m_{h}=125.5$ GeV, the production cross sections for each
production channels at LHC have been given for example in Ref.~\cite{handbook}.
In the LRTH model, the production rate of $h \to \gamma\gamma$
normalized to the SM values is generally defined as
\beq
R_{\gamma\gamma}=\frac{[\sigma(pp \to h)\times Br(h\to \gamma\gamma)]_{LRTH}}{
[\sigma(pp \to h)\times Br(h\to \gamma\gamma)]_{SM}}.
\label{eq:rggnp}
\eeq

%%%%%%%%%%%%%%%%%%%%%%%%%%%%%%%%%%%%%%%%%%%%%%%%%%
\begin{figure}[thb]
\begin{center}
\vspace{-0.5cm} \centerline{\epsfxsize=10cm\epsffile{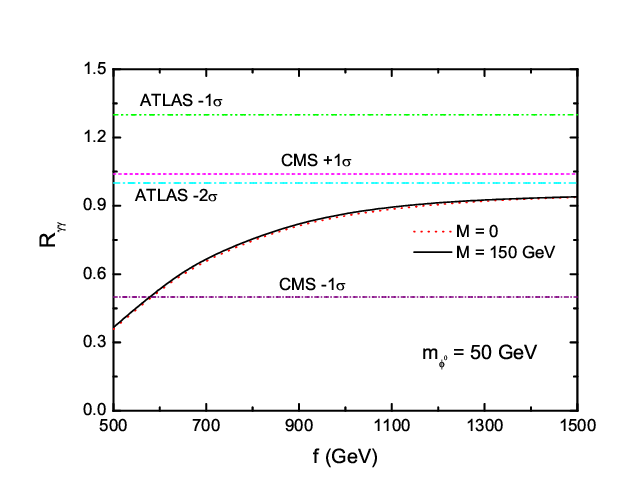}}
\caption{The $f$-dependence of $R_{\gamma\gamma}$ in the LRTH model
for $m_{\phi^0}=50$ GeV and two typical values of parameters $M$ as indicated.}
\label{fig:fig4}
\end{center}
\end{figure}

%%%%%%%%%%%%%%%%%%%%%%%%%%%%%%%%%%%%%%%%%%%%%%%%%%
\begin{figure}[tbh]
\begin{center}
%\vspace{0.0cm}
\centerline{\epsfxsize=7cm\epsffile{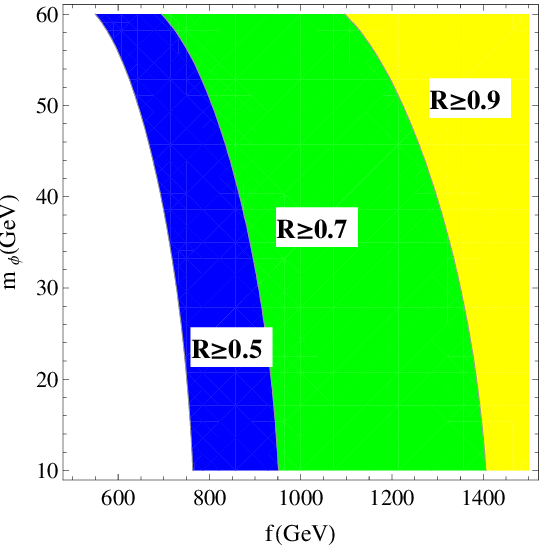}
\hspace{0.8cm}\epsfxsize=7cm\epsffile{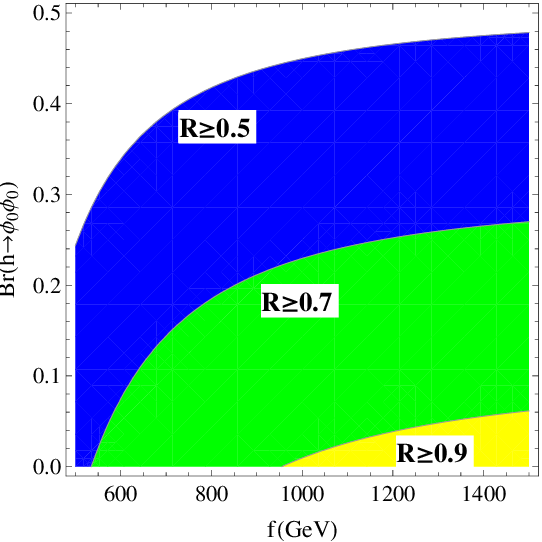}}
\caption{ The contours of $R_{\gamma\gamma}$ in $m_{\phi^0}-f$ plane (left)
and $Br(h\to\phi^0\phi^0)-f$ plane (right) for three typical
values of $R_{\gamma\gamma} \geq 0.5, 0.7$ and $0.9$. }
\label{fig:fig5}
\end{center}
\end{figure}

 In Fig.~4 we plot $R_{\gamma\gamma}$ versus $f$ for $m_{\phi^{0}}=50$ GeV and $M=0$, 150 GeV, respectively. It can be
seen from Fig.~4 that ratio $R_{\gamma\gamma}$ in the LRTH model is always smaller
than unit, and will approach one (the SM prediction) for a large $f$. On
the other hand, one can see that the ratio $R_{\gamma\gamma}$ is insensitive to the
variation of the mixing parameter $M$.
Since the ATLAS diphoton data is above the SM value by about $2\sigma$,
the predicted rate in the LRTH model is always outside the $2\sigma$ range of the ATLAS data.
But the theoretical prediction for $R_{\gamma\gamma}$ in the LRTH model
is in good agreement with the current CMS data within $1\sigma$ error for $f \geq 600$ GeV.
The key point here is the large difference between the central values reported by
ATLAS and CMS respectively.
Further improvement of the $R_{\gamma\gamma}$ measurements from both ATLAS and CMS
collaboration are greatly welcome and will play the key role in constraining
the new physics models beyond the SM.

In Fig.~5 we show the contours of $R_{\gamma\gamma}$ in
$f$-$m_{\phi^0}$ plane and $f$-$Br$ plane for
$R_{\gamma\gamma}\geq 0.5, 0.7$, and $0.9$, respectively.
One can see that the assumption $R_{\gamma\gamma}\geq$0.7 will indicate $f\geq 700$ GeV for
$m_{\phi^{0}}=60$ GeV, but leads to a limit $f\geq 900$ GeV for $m_{\phi^{0}}=30$ GeV.
From Fig.~5b, it is easy to see that one can draw strong constraint
on the exotic decay rate $Br(h\to \phi^0\phi^0 )$ from the measured Higgs diphoton rate.
A limit of $R_{\gamma\gamma} \geq$ 0.7, for example, can result in a strong
constraint $Br(h \to \phi^0\phi^0) \leq 26\%$.

\subsection{Global fit within LRTH model}

Now we perform a global fit to the LRTH model with the
method proposed in Refs.~\cite{fit1,fit11,fit2,fit21,fit3,fit4,fit41,fit,fit5}
by using the latest LHC Higgs data from both ATLAS
\cite{4,atlas,atlas1,atlas2,atlas3,atlas4,atlas5} and CMS collaboration
\cite{5,cms1,cms2,cms3,cms4,cms5}.
We use 20 sets of experimental data which include the measured signal strengths for
$\gamma\gamma$, $ZZ^{\ast}$, $WW^{\ast}$, $b\bar{b}$ and $\tau^{+}\tau^{-}$
channels, as listed explicitly in Table \ref{table1}.

\begin{table}[t,b]
\begin{center}
\caption{ The measured Higgs signal strengths $\hat{\mu}_{i}$
and the theoretical predictions $\mu_i$ in the LRTH model.
Here we set $m_{\phi^0}$=40 GeV, $M$=150 GeV and $f=800,1000,$ and $1200$ GeV.
The following corrections are included in the fit:
$\rho_{\gamma\gamma}=-0.27$, $\rho_{ZZ^{*}}=-0.5$, $\rho_{WW^{*}}=-0.18$,
$\rho_{\tau^{+}\tau^{-}}=-0.49$  for ATLAS, and
$\rho_{\gamma\gamma}=-0.5$, $\rho_{ZZ}=-0.73$ for CMS.}
\label{table1}
\vspace{0.2cm}
\begin{tabular}{|c|c|ccc|} \hline Channel & Signal strength
$\hat{\mu}_{i}$&\multicolumn{3}{c|}{LRTH predictions $\mu_i$
}\\
 & &f=800 &f=1000 &f=1200  \\ \hline
 \multicolumn{5}{|c|}{ATLAS \cite{4,atlas,atlas1,atlas2,atlas3,atlas4,atlas5}}\\
\hline
ggF+ttH, $\gamma\gamma$&$1.60\pm0.41$ &0.635&0.794&0.876 \\
\hline VBF+VH, $\gamma\gamma$&$1.94\pm0.82$ &0.726&0.856&0.928 \\
 \hline
 ggF+ttH, $ZZ^{*}$&$1.51\pm0.52$ &0.639&0.798&0.879 \\
\hline VBF+VH, $ZZ^{*}$&$1.99\pm2.12$ &0.732&0.861&0.931 \\
\hline
 ggF+ttH, $WW^{*}$&$0.79\pm0.35$  &0.639&0.798&0.879 \\
\hline VBF+VH, $WW^{*}$&$1.71\pm0.76$ &0.732&0.861&0.931 \\
\hline
 VH tag, $b\bar{b}$&$0.2^{+0.7}_{-0.6}$
 &0.720&0.861&0.931\\
 \hline
 ggF+ttH, $\tau^{+}\tau^{-}$&$2.31\pm1.61$  &0.639&0.798&0.879 \\
\hline VBF+VH, $\tau^{+}\tau^{-}$&$-0.20\pm1.06$  &0.732&0.861&0.931 \\
\hline \multicolumn{5}{|c|}{CMS \cite{5,cms1,cms2,cms3,cms4,cms5}}\\ \hline

ggF+ttH, $\gamma\gamma$&$0.49\pm0.39$ &0.635&0.794&0.876 \\
\hline VBF+VH, $\gamma\gamma$&$1.65\pm0.87$ &0.726&0.856&0.928 \\
 \hline
 ggF+ttH, $ZZ^{*}$&$0.99\pm0.46$ &0.639&0.798&0.879 \\
\hline VBF+VH, $ZZ^{*}$&$1.05\pm2.38$ &0.732&0.861&0.932 \\
\hline 0/1 jet, $WW^{*}$&$0.76\pm0.21$ &0.621&0.798&0.853 \\
\hline $Z(\nu\bar{\nu})h$, $b\bar{b}$&$1.04\pm{0.77}$ &0.720&0.861&0.925 \\
\hline $Z(l^{+}l^{-})h$, $b\bar{b}$&$0.82\pm{0.97}$ &0.720&0.861&0.925 \\
\hline $W(l\nu)h$, $b\bar{b}$&$1.11\pm{0.87}$ &0.720&0.861&0.925 \\
\hline 0/1 jet, $\tau^{+}\tau^{-}$&$0.76^{+0.49}_{-0.52}$ &0.641&0.799&0.936 \\
\hline VBF tag, $\tau^{+}\tau^{-}$&$1.40^{+0.60}_{-0.57}$ &0.734&0.873&0.936 \\
\hline VH tag, $\tau^{+}\tau^{-}$&$0.77^{+1.48}_{-1.43}$ &0.732&0.861&0.931 \\
\hline \hline $\chi^{2}$& 14.60&24.85&17.40&15.41 \\
\hline $p-$value& 0.80&0.21&0.63&0.75 \\
\hline
\end{tabular} \end{center}\end{table}

When fitting the various observable, we considered the
correlation coefficients given in Ref.~\cite{fit6} due to the
independent data for different exclusive search channels by two
collaborations. The global $\chi^{2}$ function is defined as:
\begin{equation}
\chi^{2}=\sum_{i,j}(\mu_{i}-\hat{\mu}_{i})(\sigma^{2})_{ij}(\mu_{j}-\hat{\mu}_{j}),
 \end{equation}
where $\sigma^{2}_{ij}=\sigma_{i}\rho_{ij}\sigma_{j}$,
$\hat{\mu}_{i}$ and $\sigma$ are the measured Higgs signal strengths
and their $1\sigma$ error, $\rho_{ij}$ is the correlation matrix,
$\mu_{i}$ is the corresponding theoretical predictions in terms of
the LRTH parameters. The details about the statistical treatment
are presented in Appendix A.

%%%%%%%%%%%%%%%%%%%%%%%%%%%%%%%%%%%%%%%%%%%%%%%%%%
\begin{figure}[thb]
\begin{center}
\vspace{-0.5cm}
\centerline{\epsfxsize=8cm\epsffile{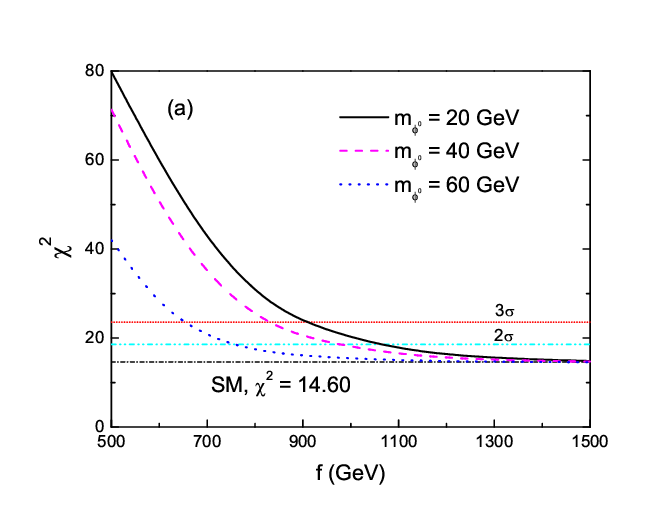}
\hspace{-0.5cm}
\epsfxsize=8cm\epsffile{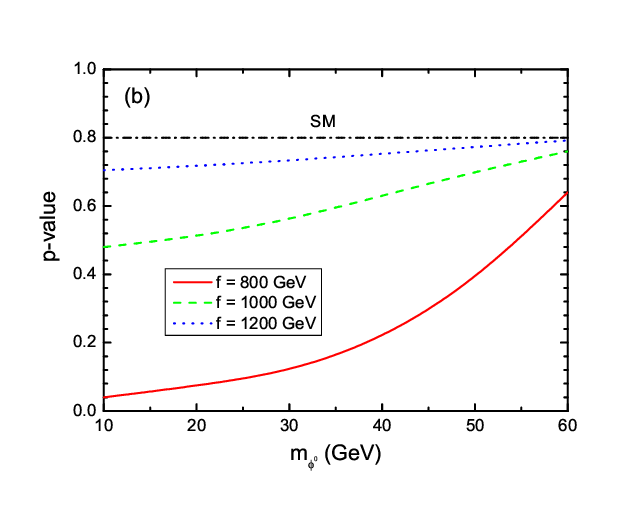}}
\caption{ (a) the values of $\chi^{2}$ versus $f$ for $M=150$ GeV and
$m_{\phi^0}=20,40$ and $60$ GeV;
(b) the $p$-values versus $m_{\phi^0}$ for $M=150$ GeV and
$f=800,1000$ and $1200$ GeV. } \label{fig:fig6}
\end{center}
\end{figure}
%%%%%%%%%%%%%%%%%%%%%%%%%%%%%%%%%%%%%%%%%%%%%%%%%

%%%%%%%%%%%%%%%%%%%%%%%%%%%%%%%%%%%%%%%%%%%%%%%%%
\begin{figure}[thb]
\begin{center}
\vspace{-1cm}
\centerline{\epsfxsize=8cm\epsffile{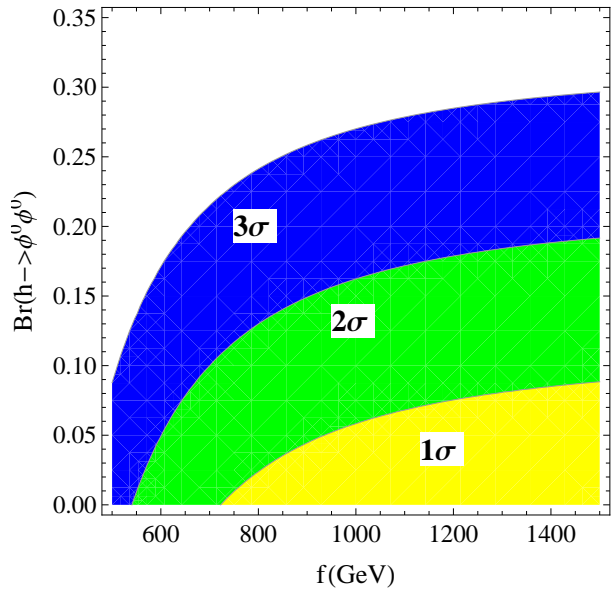}}
\caption{The contours of $\chi^{2}$ of the branching ratio $Br(h\to\phi^0\phi^0)$
at the $1\sigma$, $2\sigma$ and $3\sigma$ level. }
\label{fig:fig7}
\end{center}
\end{figure}
%%%%%%%%%%%%%%%%%%%%%%%%%%%%%%%%%%%%%%%%%%%%%%%%%

In Fig.~6a we plot $\chi^{2}$ versus $f$ for $M=150$ GeV and  $m_{\phi^0}=20,40$ and $60$ GeV,
respectively. One can see that the value of $\chi^{2}$ of the LRTH model
is larger than that for SM for most of parameter space of $f$ and approaches the SM value
for a sufficiently large $f$.
For a light pseudoscalar $\phi^0$, for example setting $m_{\phi^{0}}=20$ GeV,
the Higgs data will lead to effective constraint on the value of the parameter $f$:
$f\geq 1000$ (900) GeV at the $2\sigma$ ($3\sigma$) level.

In Fig.~6b we plot the $p$-values versus $m_{\phi^0}$ for $M=150$ GeV and
$f=800,1000$ and $1200$ GeV, respectively.
We note that the goodness of the fit in the SM, measured by the $p-$value,
is about $0.80$, which means that the SM has a chance of $80\%$ to be the true
interpretation of the data.
One can see that the $p$-value become smaller for the LRTH model in large part of its
parameter space, and approaches the SM value for a sufficiently large
$f$. For $m_{\phi^0}=40$ GeV and $f=1000 (1200)$ GeV, its $p$-value is about $0.63 (0.75)$.

In Fig.~7 we plot the contours of $\chi^{2}$ for $Br(h \to \phi^0\phi^0 )$
against the parameter $f$. One can see that the current LHC Higgs data can put strict constraint
on the exotic decay $h\to\phi^0\phi^0$: for example, $Br(h\to\phi^0\phi^0)$ should be
less than $30\%$ at 3$\sigma$ level.

\section{ Phenomenology of a light $\phi^0$}

When the decay $h\to\phi^0\phi^0$ is open, the decays $h\to\phi^0\phi^0\to 4b$, $2b2\tau$ or $4\tau$
are the major promising channels to detect such a light pseudoscalar at the LHC experiments.
As demonstrated in Ref.~\cite{prl}, the process $pp\to W/Zh\to l+4b+X$ ($l$ denotes one
lepton and $X$ denotes anything) may provide a clean signature out
of the backgrounds for a light Higgs boson. Following the suitable cuts, the signal rate
depends on an overall scaling factor
\beq
C_{4b}^{2}=\left (\frac{g_{\rm VVh}^{\rm NP}}{g_{\rm VVh}^{\rm SM}} \right )^{2}\times
Br(h\to\phi^0\phi^0)\times Br^{2}(\phi^0\to b\bar{b}),
\eeq
which determines the cross section of the process $Vh\to
V4b$ at the LHC \cite{prl,13094939}. In the LRTH model,
$y_{\rm V}=g^{\rm LRTH}_{\rm VVh}/g^{\rm SM}_{\rm VVh}=1-v^{2}/(6f^2)$ \cite{Hock}. The DELPHI
Collaboration \cite{4b} has made model-independent searches for the
process $e^{+}e^{-}\to Zh\to ZAA\to Z+4b$ with $A$ a pseudoscalar particle.
However, the experimental upper bound on $C_{4b}^{2}$ is relaxed for this
model ($C_{4b}^{2}\geq 1$ for $m_{h}=110$ GeV and $m_{A}=12$ GeV), and it is the
same case in the simplest little Higgs (SLH) model \cite{prd76}.

%%%%%%%%%%%%%%%%%%%%%%%%%%%%%%%%%%%%%%%%%%%%%%%%%%
\begin{figure}[thb]
\begin{center}
\vspace{-0.5cm}
\centerline{\epsfxsize=9cm \epsffile{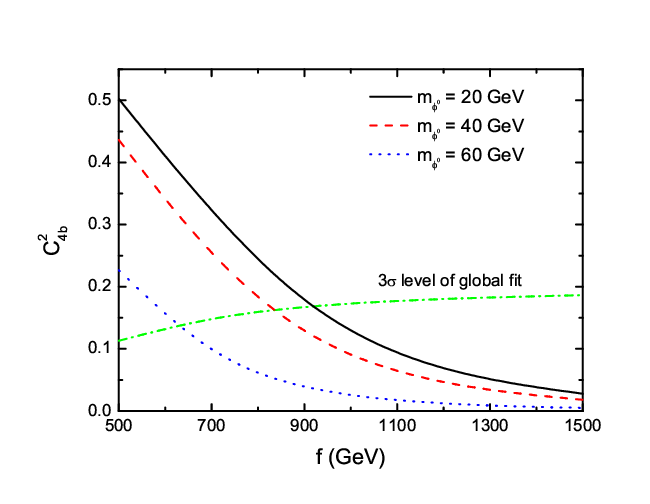}}
 \caption{ The value of $C_{4b}^{2}$ versus $f$ for
 three values of $m_{\phi^0}$.} \label{fig:fig8}
\end{center}
\end{figure}
%%=================================================

In Fig.~8 we plot the factor $C_{4b}^{2}$ versus the parameter $f$
in the LRTH model. One can see that, for $f=500$ GeV and $m_{\phi^{0}}=20$ GeV, the
value of $C_{4b}^{2}$ can be as large as 0.5. However, it is smaller than 0.2 after considering the bound of global fit
at 3$\sigma$ level.
Noticing that the value of $C_{4b}^{2}$ is directly proportional to the factor $y_{V}^{2}=(1-v^2/(6f^2) )^2$ in the LRTH model
and thus becomes larger for a large $f$.
For the process $pp\to W/Zh\to l+4b+X$, the authors of Refs.\cite{prl,13094939} have
shown that the cut on invariant mass of the four bottom quarks can suppress efficiently the relevant backgrounds.
It is worth of mentioning that Cheung et al. studied the $h\to \eta \eta $ decay \cite{prl}, calculated the total signal
and background cross sections at parton level in the SLH model with $C_{4b}^{2}=0.16$ ~\cite{prl}, and found
a significance $S/\sqrt{B}=3.7$ for a luminosity of $30$ fb$^{-1}$.
Of course, a much higher luminosity is needed to discover such a light scalar.
For example, even for $C_{4b}^{2}=0.11$ in the SLH model, the significance $S/\sqrt{B}$ can be increased from 1.4
to 4.4 for a luminosity of $300$ fb$^{-1}$. Considering the LHC Higgs data bound at $3\sigma$ level, we estimate the value of $C_{4b}^{2}$ is approximately 0.19 in the LRTH model ($C_{4b}^{2}\simeq 0.3\times 0.8^{2}$). Therefore, we hope that by using the suitable cuts, the possible signatures
of the light scalar in the LRTH model may be detected via the process $pp\to Vh\to V4b$ at the LHC with a high luminosity of $300$ fb$^{-1}$.
Certainly, detailed confirmation of the observability of the signals
would require Monte-Carlo simulations of the signals and
backgrounds, which is beyond the
scope of this paper.

%%%%%%%%%%%%%%%%%%%%%%%%%%%%%%%%%%%%%%%%%%%%%%%%%%
\begin{figure}[thb]
\begin{center}
\vspace{-0.5cm} \centerline{\epsfxsize=9cm\epsffile{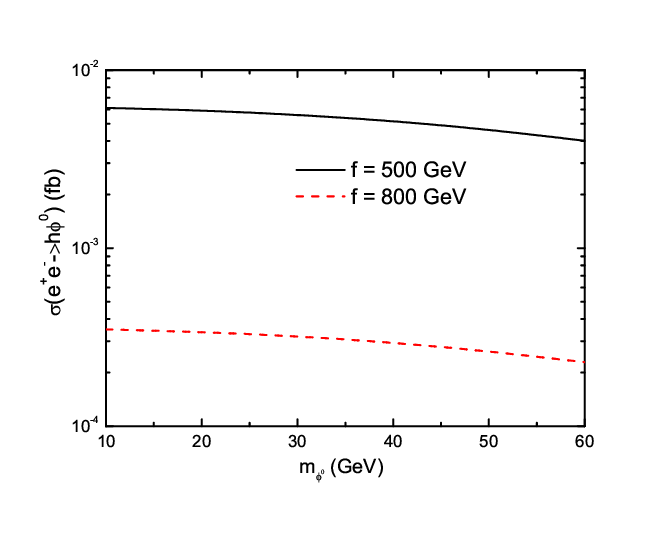}}
\caption{The cross section of $e^{+}e^{-}\to h\phi^0$ at an
electron-positron collider with $\sqrt{s}=250$ GeV for $f=500, 800$ GeV.}
\label{fig:fig9}
\end{center}
\end{figure}
%%%%%%%%%%%%%%%%%%%%%%%%%%%%%%%%%%%%%%%%%%%%%%%%%%
\begin{figure}[tbh]
\begin{center}
\vspace{-0.5cm} \centerline{\epsfxsize=8cm\epsffile{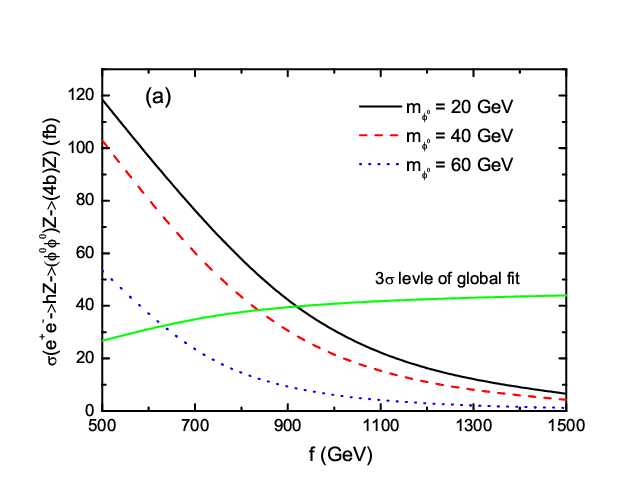}
\hspace{-0.5cm}\epsfxsize=8cm\epsffile{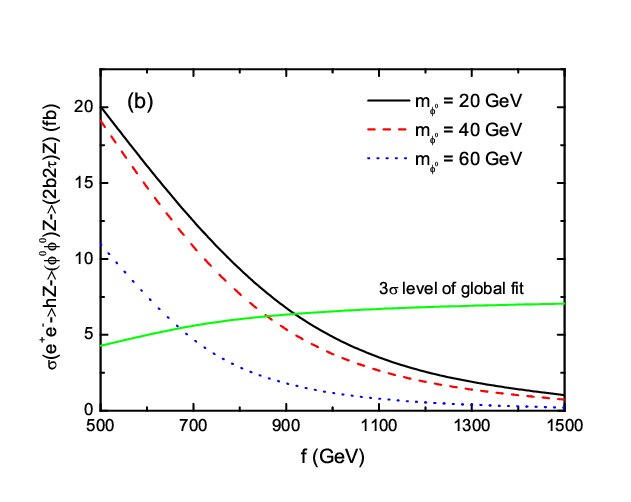}}
\caption{ The cross sections at an electron-positron collider with $\sqrt{s}=250$ GeV
and $m_{\phi^0}=(40\pm 20)$ GeV; (a) $e^{+}e^{-}\to Zh\to Z(\phi^0\phi^0)\to Z(4b)$,
(b) $e^{+}e^{-}\to Z h\to Z(\phi^0\phi^0)\to Z(2b2\tau)$. }
\label{fig:fig10}
\end{center}
\end{figure}

The light scalar $\phi^0$ can also be produced associated with the SM-like Higgs $h$ at the International Linear Collider (ILC),
which has been studied in Ref.~\cite{hh}.
The numerical results show that the resonance production cross section
can be significantly enhanced at the high energy linear collider with $\sqrt{s}\simeq
m_{Z_{H}}$. On the other hand, the properties of SM-like Higgs $h$ can
be precisely measured through the $Zh$ associated production at the linear collider \cite{zh,zh1,zh2}.
Here we calculate the cross sections of the process $e^{+}e^{-}\to h\phi^0$ and
$e^{+}e^{-}\to Zh\to Z(\phi^0\phi^0)\to Z(4b,2b2\tau)$ at an electron-positron collider with $\sqrt{s}=250$
GeV, as shown in the Fig.~10.
As shown in Fig.~9, the associated production rate $Br(e^+e^-\to h \phi^0)$ is smaller than
the order of $10^{-2}$ fb at $\sqrt{s}$=250 GeV, which can hardly be utilized to search for the light scalar $h$.
However, the production cross sections of processes $e^{+}e^{-}\to Zh\to Z4b$ and $e^{+}e^{-}\to Zh\to Z(\phi^0\phi^0)\to
Z(2b2\tau)$ can reach 120 fb and 20 fb respectively, as illustrated in Fig. 10.
Certainly, the cross sections would become smaller when we consider the global fit bound
at $3\sigma$ level (reduced about two thirds). Since these signals are free of the SM background,
such production process may contribute the light scalar discovery at an electron-positron collider.

\section{ Conclusions}

The LRTH model predicts one neutral pseudoscalar particle $\phi^0$, which may be lighter than half of the Higgs boson
mass. In this work we focus on the case of $m_{\phi^0} < m_h/2$ so that the new decay mode $h\to \phi^0\phi^0$
can be open. In this work, we firstly calculated the decay widths and the branching ratios of
the $h \to \phi^0 \phi^0$ decay , as well as the major decay modes of the $\phi^0$ itself: such as
$\phi^0\to (b\bar{b}, c\bar{c}, \tau^+\tau^-)$ and $ \phi^0\to( gg, \gamma\gamma)$ decays.
We then examined the $f$, $M$ and $m_{\phi^0}$-dependence of the decay widths and corresponding branching ratios,
and checked the possible constraints on the LRTH model from the latest LHC Higgs data on such a possibility.
We performed a global fit by using 20 sets of the measured Higgs signal strengths as reported by ATLAS and CMS
collaboration for $\gamma\gamma$, $ZZ^{\ast}$, $WW^{\ast}$, $b\bar{b}$ and $\tau^{+}\tau^{-}$ channels.
We also studied the detection of $\phi^0$ at future electron-positron collider experiments.

From our numerical calculations and the phenomenological analysis we found the following points:
\begin{enumerate}
\item
Without the LHC constrains, the branching ratio of the decay $h\to \phi^0\phi^0$ can be as large as
$80\%$ and it can suppress significantly the visible $\gamma\gamma$ signal rate.
The current LHC Higgs data for the $\gamma\gamma$ channel can place strong limit
on such a decay: for example, $Br(h\to \phi^0\phi^0) \leq 26\%$ for $R_{\gamma\gamma}\geq 0.7$.

\item
The $p$-value of the SM Higgs boson is $0.80$, which means that the SM is a reasonably good fit to the Higgs data.
In the LRTH model, its $p$-value is smaller than that of the SM in most of the parameter
space  and approaches the SM value for a sufficiently large $f$ parameter.

\item
The latest LHC Higgs data constrain the branching ratio $Br(h \to \phi^0\phi^0)$ to be less than $30\%$ at 3$\sigma$ level.

\item
The neutral scalar $\phi^0$ dominantly decay into $b\bar{b}$ and the decay rate $Br(\phi^0\to b\bar{b})$ can be
larger than $80\%$ for $m_{\phi^0}\leq 60$ GeV.
The second main decay mode is $\phi^0\to \tau^{+}\tau^{-}$ with a branching ratio about $14\%$.
At the future $e^--e^+$ collider with $\sqrt{s}=250$ GeV, the processes $e^{+}e^{-}\to Zh\to
Z(\phi^0\phi^0)\to Z(4b,2b2\tau)$ are promising for discovering such a light pseudoscalar $\phi^0$.

\end{enumerate}

\begin{acknowledgments}

We thank Shufang Su for providing the Calchep Model Code.
This work is supported by the National Natural Science
Foundation of China under the Grant No. 11235005 and the Joint Funds
of the National Natural Science Foundation of China (U1304112).

\end{acknowledgments}

%%%%%%%%%%%%%%%%%%%%%%%%%%%%%%%%%%%%%%%%%%%%%%%%%%%%%%%%%%%%%%%%%%%%%%%%%%%%%%%%%%
%                                        Appendix
%%%%%%%%%%%%%%%%%%%%%%%%%%%%%%%%%%%%%%%%%%%%%%%%%%%%%%%%%%%%%%%%%%%%%%%%%%%%%%%%5

\begin{appendix}

\section{The statistical treatment and data}\label{sec:data}

Take the $h\to \gamma\gamma$ for instance, the Higgs signal strength $\mu_{\gamma\gamma}$ can be defined as
\beq
\mu_{\gamma\gamma}=\frac{\epsilon_{ggF}\sigma_{ggF}+\epsilon_{VBF}
\sigma_{VBF}+\epsilon_{VH}\sigma_{VH}}{\epsilon_{ggF}\sigma^{SM}_{ggF}
+\epsilon_{VBF}\sigma^{SM}_{VBF}+\epsilon_{VH}\sigma^{SM}_{VH}}\times\frac{Br(h\to
\gamma\gamma)}{Br(h\to \gamma\gamma)_{SM}},
\eeq
where the coefficients $\epsilon$ accounting for the relative weight
of each production channel given in \cite{4,5,fit2}. The SM production
cross sections and decay widths are taken from \cite{handbook}.

The errors on the reported Higgs signal strengths $\hat{\mu}_{i}$ are symmetrized by
\beq
\delta\hat{\mu}_{i}=\sqrt{\frac{(\delta\hat{\mu}_{+})^{2}+(\delta\hat{\mu}_{-})^{2}}{2}},
\eeq
where $\delta\hat{\mu}_{\pm}$ are the one-sided errors given by the
experimental collaborations \cite{4,5}.
For plotting distributions of a function of one (two) parameter, the $68\%$ ($1\sigma$), $95\%$
($2\sigma$) and $99.7\%$ ($3\sigma$) confidence level (CL) intervals are obtained by
$\chi^{2}=\chi_{min}^{2}+$1 (2.3), + 4 (6.18), and +9 (11.83), respectively \cite{fit4}.

For two correlated observables, the correlation coefficient $\rho$ is applicable to the
following formula
\beq
\chi_{1,2}^{2}=\frac{1}{(1-\rho^{2})}\cdot[\frac{[\mu_{1}-\hat{\mu}_{1}]^{2}}{\sigma_{1}^{2}}
+\frac{[\mu_{2}-\hat{\mu}_{2}]^{2}}{\sigma_{2}^{2}}-2\rho\frac{[\mu_{1}-\hat{\mu}_{1}]
\cdot[\mu_{2}-\hat{\mu}_{2}]}{\sigma_{1}\sigma_{2}}].
\eeq

Assuming the goodness-of-fit statistics follow a $\chi^{2}$ probability density function,
the $p-$value for the hypothesis is given by \cite{data}
\beq
p=\int_{\chi^{2}}^{\infty}\frac{z^{n/2-1}e^{-z/2}}{2^{n/2}\Gamma(n/2)}dz,
\eeq
where $n$ is the degrees of freedom ($n=20$ in this work).
\end{appendix}

%%%%%%%%%%%%%%%%%%%%%%%%%%%%%%%%%%%%%%%%%%%%%%%%%%%%%%%%%%%%%%%%%%%%%%%%%%%%%%%%%%%%%%%%%%%%%%5
%                                 reference
%%%%%%%%%%%%%%%%%%%%%%%%%%%%%%%%%%%%%%%%%%%%%%%%%%%%%%%%%%%%%%%%%%%%%%%%%%%%%%%%%%%%%%%%%%%%%%%%%

\end{document}